\documentclass[a4paper,fleqn]{cas-dc}
\usepackage[square,numbers]{natbib}
\begin{document}
\let\WriteBookmarks\relax
\def\floatpagepagefraction{1}
\def\textpagefraction{.001}

\shorttitle{Quantifying energy fluence and its uncertainty for radio emission}

\shortauthors{S. Martinelli et~al.}

\title [mode = title]{Quantifying energy fluence and its uncertainty for radio emission from particle cascades in the presence of noise}                      

\author[1,2]{Sara Martinelli}[type=editor,
                        auid=000,bioid=1,
                        orcid=0009-0004-6927-6301]

\cormark[1]
\ead{sara.martinelli@kit.edu}

\affiliation[1]{organization={Institute for Astroparticle Physics, Karlsruhe Institute of Technology (KIT)},
    city={Karlsruhe},
    country={Germany}}
    
\affiliation[2]{organization={Instituto de Tecnologías en Detección y Astropartículas (CNEA, CONICET, UNSAM)},
    city={Buenos Aires},
    country={Argentina}}  
    
\author[1,3]{Tim Huege}[type=editor,
                        auid=000,bioid=1,
                        orcid=0000-0002-2783-4772]

\ead{tim.huege@kit.edu}

\affiliation[3]{organization={Astrophysical Institute, Vrije Universiteit Brussel}, 
    city={Brussels},
    country={Belgium}}

\author[2]{Diego Ravignani}[type=editor,
                        auid=000,bioid=1,
                        orcid=0000-0001-7410-8522]
\ead{diego.ravignani@iteda.cnea.gov.ar}

\author[4,5]{Harm Schoorlemmer}[type=editor,
                        auid=000,bioid=1,
                        orcid=0000-0002-8999-9249]
\ead{hschoorlemmer@science.ru.nl}

\affiliation[4]{organization={IMAPP, Radboud University Nijmegen}, 
    city={Nijmegen},
    country={The Netherlands}}
    
\affiliation[5]{organization={Nationaal Instituut voor Kernfysica en Hoge Energie Fysica (NIKHEF)}, 
    city={Amsterdam},
    country={The Netherlands}}

\cortext[cor1]{Corresponding author}

\begin{abstract}
Measurements of radio signals induced by an astroparticle generating a cascade present a challenge because they are always superposed with an irreducible noise contribution. Quantifying these signals constitutes a non-trivial task, especially at low signal-to-noise ratios (SNR). Because of the randomness of the noise phase, the measurements can be either a constructive or a destructive superposition of signal and noise. To recover the electromagnetic energy of the cascade from the radio measurements, the energy fluence, i.e. the time integral of the Poynting vector, has to be estimated. Conventionally, noise subtraction in the time domain has been employed for energy fluence reconstruction, yielding significant biases at low signal-to-noise ratios. In several analyses, this bias is mitigated by imposing an SNR threshold cut, though this option is not ideal as it excludes valuable data. Additionally, the uncertainties derived from the conventional method are underestimated, even for large SNR values. To address this known issue, the uncertainties have so far typically been approximated and corrected by using ad-hoc terms. This work tackles these challenges by detailing a method to correctly estimate the uncertainties and lower the reconstruction bias in quantifying radio signals, thereby, ideally, eliminating the need for an SNR cut. The development of the method is based on a robust theoretical and statistical background, and the estimation of the fluence is performed in the frequency domain, allowing for the improvement of further analyses by providing access to frequency-dependent fluence estimation.
\end{abstract}


\maketitle

\section{Introduction}
\label{sec:Introduction} 
Radio detection can be employed to measure cosmic-rays, photons, and neutrinos generating extensive air showers and particle cascades in dense media \cite{Tim_review, Frank_review}. Since the radio emission is proportional to the number of electrons and positrons in the cascade, by quantifying the underlying signal of a measured radio pulse we can access the electromagnetic component of a shower, and, thus, estimate the electromagnetic energy of the originating particle.  As the electromagnetic energy is proportional to the area integral of the energy fluence (the energy deposit per unit area in terms of radio waves) \cite{AERA_Energy_PRL}, a correct reconstruction of the electromagnetic energy and its uncertainty relies on the determination of the energy fluence and its uncertainty. 

The total energy fluence $f_\mathrm{tot}$ at a given antenna position is the time integral of the Poynting vector. For discretely sampled measurements, this means:
\begin{equation}
f_\mathrm{tot}= \epsilon_0\,c\,\Delta t \sum_\mathrm{pol} \bigg(\sum_{j} E^2_\mathrm{pol}(t_j) \bigg) , 
\label{eq:poynting}
\end{equation}
where $\Delta t$ is the sampling interval of $\Vec{E}(t)$, the observed three-dimensional electric-field vector (in the equation broken down into its components), $\epsilon_0$ is the vacuum permittivity, and $c$ is the speed of light. In an analogous way, we can express the energy fluence as the sum of component-dependent contributions $f_\mathrm{pol}$:
\begin{equation}
f_\mathrm{pol} =\epsilon_0\,c\,\Delta t \sum_{j} E^2_\mathrm{pol}(t_j) \,\, \rightarrow \,\, f_\mathrm{tot}= \sum_\mathrm{pol} f_\mathrm{pol}. 
\label{eq:poynting_component}
\end{equation}
In radio measurements, the noise requires a sophisticated treatment: the measured pulse is either enhanced or diminished compared to the true signal depending on the random phase of the noise. This makes the evaluation of the fluence in the presence of noise non-trivial, especially at low signal-to-noise ratios (SNR) \cite{Frank_noise}.\\

So far, the conventional way of reconstructing the signal energy fluence consisted of estimating and subtracting the noise fluence in the time domain \cite{Glaser_thesis}. This way of treating the noise leads to a non-negligible reconstruction bias at low SNR values. To mitigate potential reconstruction bias, especially in the case of electromagnetic energy, a minimum required SNR value can be introduced as a prerequisite to using data from a given antenna. This approach is used, for example, in references \cite{xmx_PRD, MarvinARENA2024, MaxEnergyScaleARENA2024}, where a cut is applied at the electric field level for inclusion of a given antenna station. One of the aims of this work is to diminish the reconstruction bias allowing the lowering and, ideally, removal of the need for an SNR cut. Furthermore, the uncertainties derived from the noise-subtraction method are known to be underestimated \cite{AERA_energy_PRD, Felix_thesis, MarvinARENA2024}. In this work, we detail a method for the quantification of the energy fluence and its uncertainty, exploiting a solid statistical background based on Rice distributions. The same statistical background has already been used within the ANITA experiment to achieve an estimation of the fluence in the frequency domain by employing fitting procedures \cite{Schoorlemmer_2016}. Our method is based on estimating the fluence in the frequency domain, too, i.e., providing access to a frequency-dependent fluence estimation. Since the spectral shape can be described more easily than the pulse shape in the time domain, the method presented here can be further developed by combining it with the information derived from spectral modeling \cite{Nikos_ICRC2023}. In the following, we validate the Rice-distribution method and compare it to the so-far widely adopted noise subtraction method. The bias of the reconstructed fluence as a function of the SNR will be discussed, as well as the evaluation of the uncertainties. 
\begin{figure*}[h!]
\centering
\includegraphics[width=0.9\linewidth]{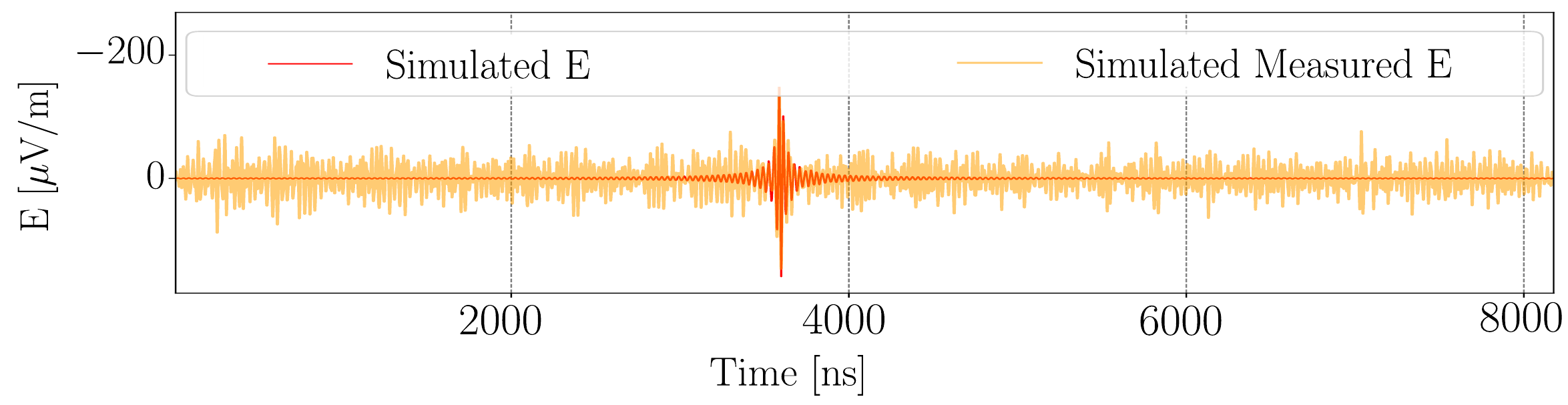}
\caption{Example of an electric-field component simulated in the absence of noise (red) and in the presence of noise (orange).}
\label{fig:Efield_exampla}
\end{figure*}

In the following, both methods are applied to the same data set of simulated air showers initiated by cosmic-rays. Because of its complexity, it is a non-trivial task to artificially generate realistic noise. To realistically simulate radio measurements, we have added the ambient background recorded at a particular site of the Pierre Auger Observatory to the radio-emission simulations. By using these measurements we get a realistic SNR distribution. We simulate the antenna response of the Radio Detector (RD) \cite{TimUHECR2022} of the Pierre Auger Observatory, sensitive to the 30-80\,MHz frequency bandwidth. Nevertheless, the results of our study are independent of the detector simulation, and the Rice-distribution method can be applied to larger frequency bandwidths than the one tested in this work. 

In the Appendix \ref{sec:s_ML_estimator}, we also present a method for estimating the spectral amplitudes of the signal and their uncertainties based on maximizing the Rice likelihood function. In the Appendix \ref{appendix:table}, the reader can find tables listing most of the notation and variables adopted in the following.

\section{Simulated Data Set} \label{Data-set}
Here, we shortly describe the set of simulations used to study the performance of the estimation methods.
\subsection{Particle Simulations and Simulated Radio-footprints } 
We exploit the same set of simulated air showers as used in reference \cite{FelixICRC2021}. The showers are initiated by four different cosmic-ray primaries: proton, helium, nitrogen, and iron. For each primary, the particle and radio-emission footprints of about 2000 showers are simulated with CORSIKA/CoREAS v7.7401 \cite{CoREAS, CORSIKA}, using the high energy interaction model QGSJETII-04 \cite{QGSJET} and an optimized thinning level of $10^{-6}$ \cite{thinning}. The primaries have energies in the interval between $10^{18.4}$\,eV and $10^{20.1}$\,eV. The arrival directions of the showers are uniformly distributed across azimuth angles, while the zenith angle $\Theta$, ranging from $65^\circ$ to $85^\circ$, is uniformly distributed according to a $\mathrm{cos}^2(\Theta)$ weighting. The environmental conditions are set to match the Pierre Auger Observatory site, as is the detector layout. The cores of the showers are randomly distributed within this detector layout. 

\subsection{Simulations of Electric Fields in the Absence of Noise}\label{efield_sim}
The CoREAS simulations are processed through the reconstruction framework of the Pierre Auger Observatory, $\overline{\mathrm{Off}}\underline{\mathrm{line}}$ \cite{Offline_radio}. The RD response simulation is performed and the electric field traces are reconstructed from the raw signal traces by using the Monte Carlo values of the arrival direction and the sensitivity pattern of the RD antennas. This means that the antenna response is applied to the CoREAS radio pulses and then again deconvolved. The electric field traces obtained correspond to the air shower pulses as they would be reconstructed in the absence of noise, in the 30-80\,MHz frequency band. Each trace has a length of 8192\,ns and $\Delta t$\,$=$\,1\,ns, upsampled from the nominal 250 MSPS sampling rate with a factor of four. 

\subsection{Simulations of Measured Electric Fields in the Presence of Noise}\label{efield_meas} 
To get the simulated measurements of the cosmic-ray pulses in the presence of noise, we apply the detector simulation to the CoREAS simulations and add measured noise data. The measured traces of the ambient background recorded at the Pierre Auger Observatory site operating over one year are added to the digitized simulated traces. Each noise measurement is used multiple times by rolling the trace. In figure \ref{fig:Efield_exampla}, we show an example of a simulated trace in the absence of noise and the corresponding trace simulated in the presence of noise, obtained as just described.  

\section{The Noise Subtraction Method}\label{Offline_method}
\begin{figure*}[b!]
\centering
\includegraphics[width=0.98\linewidth]{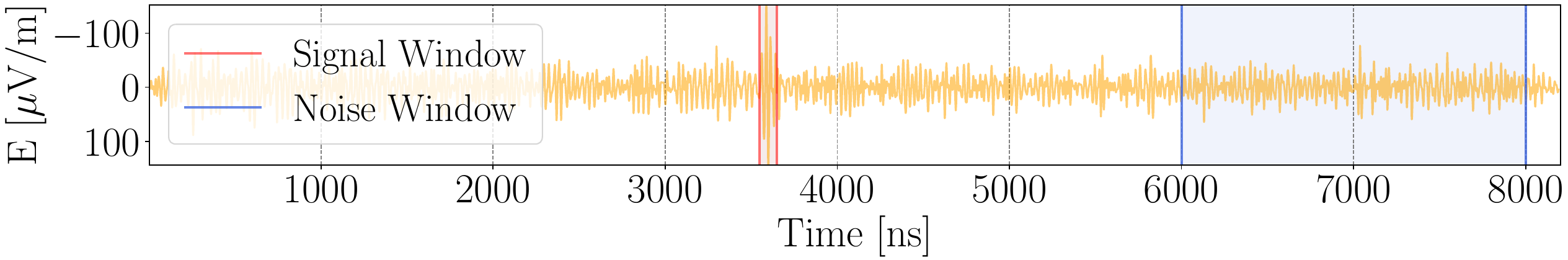} 
\caption{
Signal window (between red lines) and noise window (between blue lines) as used in the noise subtraction method.}
\label{fig:offline_windows}
\end{figure*}
The noise subtraction method is based on the assumption that the measured amplitude of the electric field is given by the sum of the cosmic-ray radio pulse and Gaussian-distributed noise $e(t)$, having a certain standard deviation $\sigma_e$ and centered on zero \cite{Glaser_thesis}. To estimate the energy fluence, we define a \textbf{noise window}, delimited by $t_1$ and $t_2$, and a \textbf{signal window}, delimited by $t_3$\,$=$\,$t_\mathrm{peak}$-$\,\Delta$ and $t_4$\,$=$\,$t_\mathrm{peak}$+\,$\Delta$, where $t_\mathrm{peak}$ indicates the position of the radio pulse in the time trace, and $\Delta$ has to be chosen such as to contain most of the pulse. As shown in figure \ref{fig:offline_windows}, the noise window has to be far away from the signal window to contain mainly only noise contribution. For each component of the measured electric field at a given antenna position $E_\mathrm{pol}(t)$, the energy fluence is estimated as:
\begin{equation}
\hat{f}_\mathrm{pol}=\epsilon_0\,c\,\Delta t \bigg (\sum_{t_j=t_3}^{t_4}E_\mathrm{pol}^2(t_j) - \frac{t_4 -t_3 }{t_2-t_1}\sum_{t_j=t_1}^{t_2} E_\mathrm{pol}^2(t_j)\bigg ), 
\label{eq:noise_sub}
\end{equation}
where the normalized fluence in the noise window is subtracted from the fluence calculated in the signal window. When dealing with very noisy pulses, the above equation can yield negative values. To prevent unphysical results, negative-valued estimators are typically set to zero, and we do the same in the analysis presented here. The uncertainty on the fluence estimator, as derived in \cite{Glaser_thesis}, consists of:
\begin{equation}
\delta\big(\hat{f}_\mathrm{pol}\big)=\sqrt{
4\,\epsilon_0\, c\, \Delta t\,\hat{f}_\mathrm{pol}\,\sigma_e^2\, +\, 2\,(\epsilon_0\,c)^2\,\Delta t\,\sigma_e^4},
\end{equation}
where $\sigma_e$ can be approximated with the root-mean-square of the trace computed in the noise window. The estimator of the total energy fluence at the antenna position is given by $\hat{f}_\mathrm{tot}$=$ \sum_\mathrm{pol} \hat{f}_\mathrm{pol}$, and its uncertainty is obtained by propagating the errors $\delta\big(\hat{f}_\mathrm{pol}\big)$.

\begin{figure*}[h!]
\centering
\includegraphics[width=0.98\linewidth]{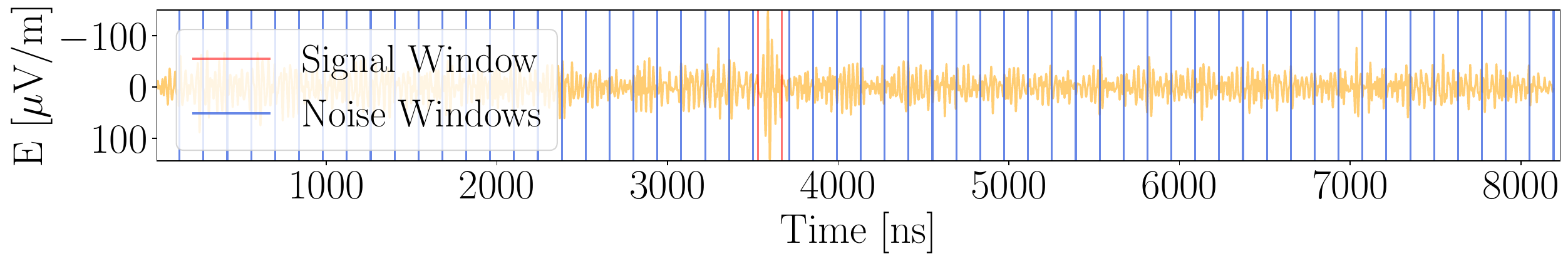} 
\caption{Signal window (between red lines) and noise windows (between blue lines) as used in the Rice-distribution method. In the example, $N$=57 noise windows are used. }
\label{fig:rice_allwindows}
\end{figure*}
\begin{figure*}[t!]
\centering
\includegraphics[width=1\linewidth]{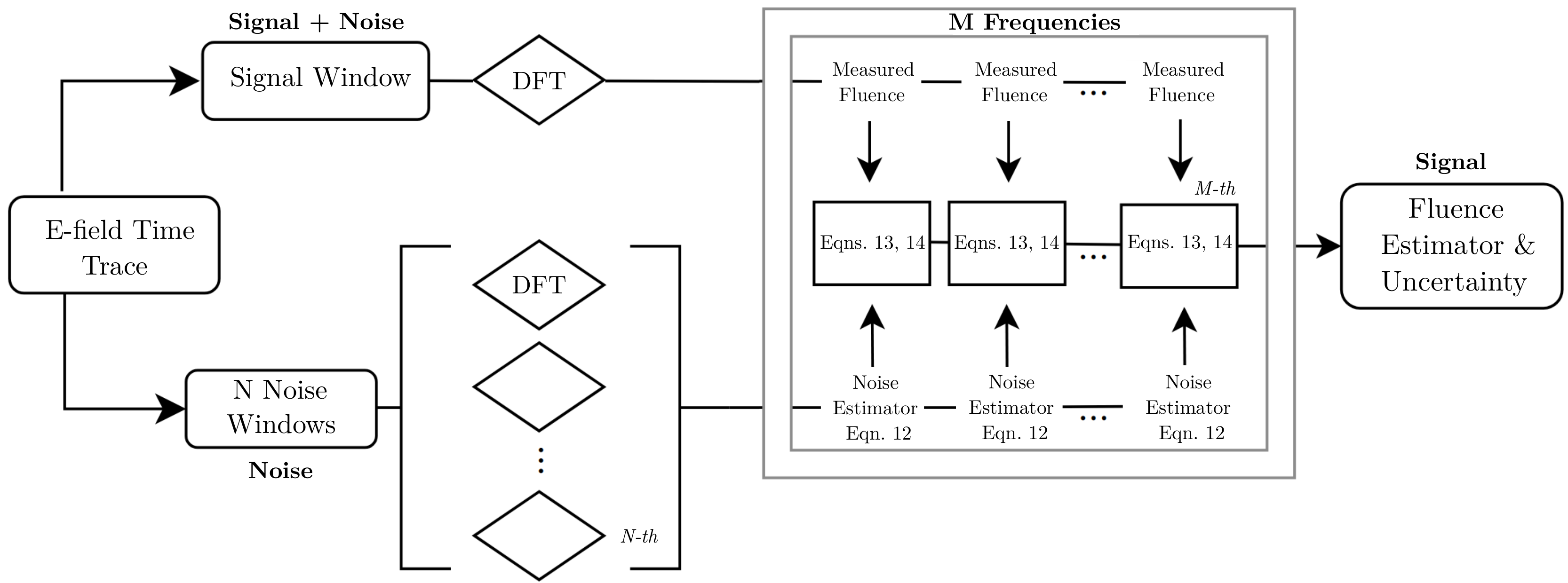}
\caption{Flowchart representing the logic of the energy fluence estimation for a single polarisation of the electric field. Combining the $M$ estimators of the noise fluence obtained from the $N$ noise windows, and the fluence measured in the signal window, we get the polarisation fluence estimator and its uncertainty.}
\label{fig:flowchart}
\end{figure*}

\section{Rice-distribution Method}
Unlike the relatively straightforward noise subtraction method, the Rice-distribution method comprises several steps. Here, we summarize the general logic behind the method, with detailed explanations and discussion provided in the subsequent sections\footnote{The implementation of the algorithm can be found at https://gitlab.iap.kit.edu/saramartinelli/fluence-and-uncertainty-estimation-based-on-rice-distribution.}.

For each component of the measured electric field at a given antenna position, we define a signal window and $N$ noise windows along the time trace (see figure \ref{fig:rice_allwindows}). Since our method relies on estimating the energy fluence in the frequency domain, we ensure a correct and efficient evaluation by applying \textit{windowing} before performing any Fourier transform (sec.\ \ref{sec:rice_windows}). This results in a signal-window frequency spectrum of $M$ Rice-distributed spectral amplitudes, and $N$ noise frequency spectra, each having $M$ spectral amplitudes assumed to be Rayleigh-distributed  (sec.\ \ref{sec:rice_theo}). We express the measured fluence as the sum of $M$ frequency-dependent contributions. For each frequency, we estimate the noise fluence over the $N$ windows and, finally, estimate the signal fluence and its uncertainty. A summary of the formulas used is provided in section \ref{sec:fluence_nck2}, while the derivation of these formulas -  obtained by exploiting the statistical background based on the Rice distribution -  can be found in Appendix \ref{sec:derivation_fluence_nck2}. In sec.\ \ref{sec:toy}, we study the bias of the signal fluence estimator and the coverage of its error by performing a toy Monte Carlo for a single frequency. 
Finally, by summing up the $M$ signal estimators and propagating the errors, we obtain the estimator of the polarization fluence and its uncertainty. The logic is illustrated in the flowchart of figure \ref{fig:flowchart}. 

To finally estimate the total fluence at the antenna position and its uncertainty, we repeat this process for the remaining polarizations.

\subsection{Signal and Noise Windows}
\label{sec:rice_windows}

When calculating a discrete Fourier transform (DFT), we implicitly assume that the finite sequence of considered samples is periodic in time. If the start and end points of the sequence do not match each other, artificial spectral contributions are introduced in the frequency spectrum. To reduce the spectral leakage, it is recommended to apply a tapering function \cite{Windows}. In the implementation of our method, anytime we perform a DFT of a sequence of samples, we employ a fast Fourier transform (FFT) algorithm, and first we apply in the time domain the so-called \textit{Tukey} window \cite{Tukey}. This is a symmetrical function consisting of a rectangular function combined with two halves of a \textit{Hann} window, also known as \textit{raised cosine bell} window \cite{Windows}. For this reason, the Tukey window is often referred to as \textit{split cosine bell} or \textit{cosine-tapered} window. Given a Tukey window of total length $L_\mathrm{tot}$\,=\,$N_\mathrm{tot}\Delta\,t$, where $N_\mathrm{tot}$ is the total number of the samples of the considered sequence, one can require the proportion $p$ of the sequence to be tapered by the two halves of the Hann window. In other words:
\begin{equation}
L_\mathrm{H}=p\,L_\mathrm{tot}=N_\mathrm{H}\,\Delta t, \quad \mathrm{with}\,\,\,\,N_\mathrm{H}=p\,N_\mathrm{tot},
\label{eq:Tukey}
\end{equation}
where $L_\mathrm{H}$ is the length of the Hann window and $N_\mathrm{H}$ is the total number of samples covered by its two halves. 

\begin{figure*}[h!]
\centering
\includegraphics[width=0.99\linewidth]{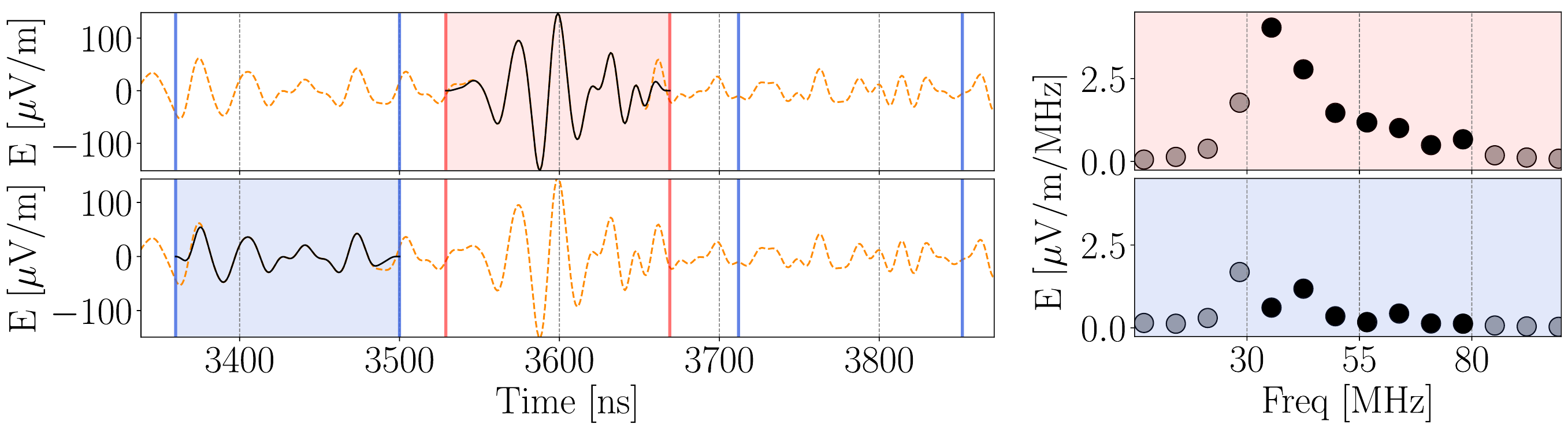} 
\caption{Examples of signal window (upper plots) and noise window (lower plots) as defined for the Rice-distribution method. On the left, we show the time trace before (orange) and after (black) applying the Tukey function in the signal window (red) and the noise window (blue). On the right, the corresponding frequency spectra are shown using the same color code, with the black dots indicating the spectral amplitudes included in the 30-80\,MHz frequency band.}
\label{fig:rice_wind_spectra}
\end{figure*}

To properly define the \textbf{signal window}, we employ the Tukey window just introduced. First, we fix $L_\mathrm{tot}$\,$=$\,140\,ns, and $L_\mathrm{H}$\,$=$\,40\,ns, then, we clip the time trace around the pulse position $t_\mathrm{peak}$ such that the resulting clipped trace has a length equal to $L_\mathrm{tot}$. Finally, we apply the Tukey window and we perform the FFT of the windowed trace. Because of the usage of windowing, the time trace is damped at the edges of the signal window, where usually the noise is stronger than the signal (see figure \ref{fig:rice_wind_spectra}). The resulting spectrum could present a non-negligible contribution outside the sensitive frequency band of interest, mainly due to the noise. For this reason, these spectral amplitudes are not considered. To evaluate the noise level of the measurement, we now define \textbf{$N$ noise windows}. Starting from the beginning of the trace, we apply the Tukey window every multiple of $L_\mathrm{tot}$, until we cover the entire trace length. We take care of skipping the signal window and applying additional spacing to reduce the signal contribution to the noise evaluation. In this work, we use a spacing of $\pm$\,20\,ns and the same $L_\mathrm{tot}$ and $L_\mathrm{H}$ employed for the signal window. This results in having about $N$\,$\approx$\,60 noise windows per measured trace. We perform the FFT in each noise window and we exclude the spectral amplitudes outside the sensitive frequency band of interest, as done in the signal window. As shown in figure  \ref{fig:rice_wind_spectra}, we get $M$\,$=$\,7 spectral amplitudes within the 30-80\,MHz bandwidth both for the signal window and each of the noise windows.

\begin{figure*}[h!]
\centering
\includegraphics[width=0.89\linewidth]{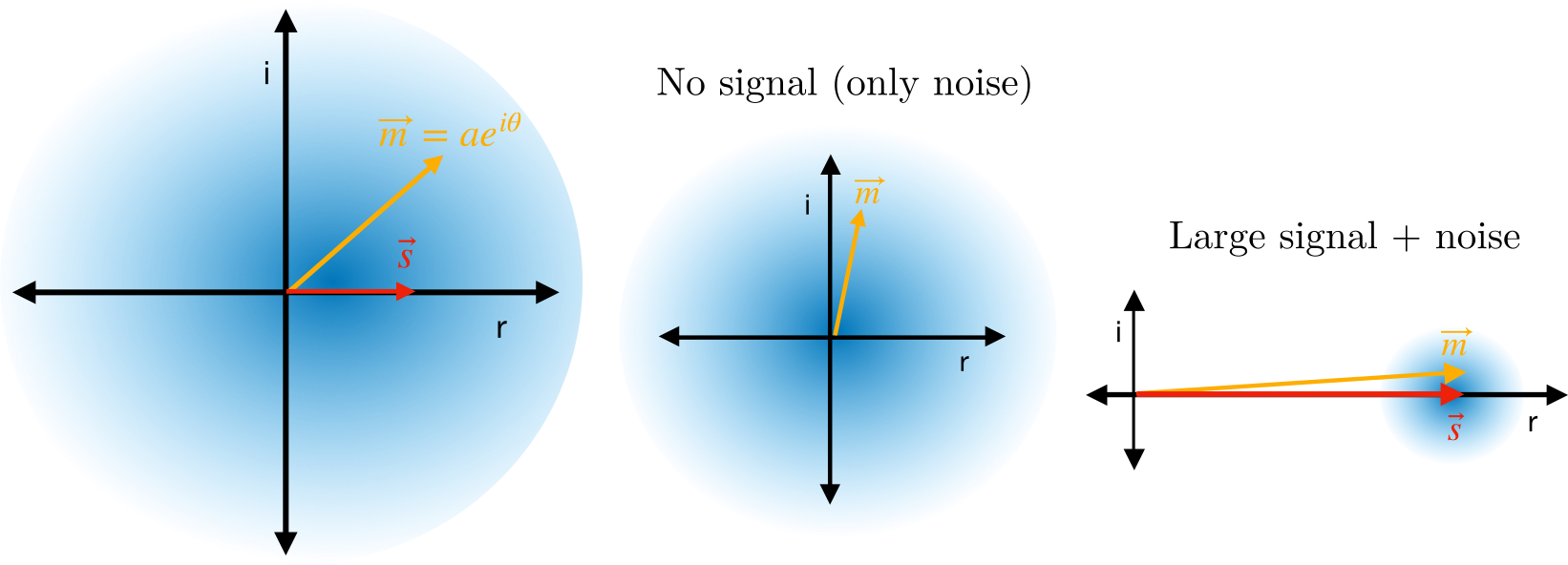}
\caption{Representation of radio measurements using the phasor formalism. In the images, the signal phasor $\Vec{s}$ is depicted on the real axis (red). The measurement $\Vec{m}$ (orange) is given by the sum of $\Vec{s}$ and the sum of the $\mathcal{N}$ random noise phasors. The probability density function of the noise phasor is illustrated as a blue cloud. }
\label{fig:phasors}
\end{figure*}

\subsection{Rice Distribution}
\label{sec:rice_theo}
After defining the signal window and performing the FFT, we obtain $M$ measured spectral amplitudes $a(\nu_j)$, where the index $j$ runs over $M$. These amplitudes are given by the superposition of the true signal and the measurement noise of each frequency bin. To recover the unknown spectral amplitudes of the true signal $s(\nu_j)$, we need to use a formalism that allows us to properly disentangle the phase information from the amplitude. To do this, the phasors formalism can be adopted. 

Let us first focus on a single-frequency bin, where $\Vec{s}$ is a constant phasor representing the true signal, and $\Vec{m}$\,=\,$a\,e^{i\theta}$ is the measurement phasor, having amplitude $a$ and phase $\theta$. Following the derivation of reference \cite{StatOptics}, we express the noise contribution to the measurement as the sum of $\mathcal{N}$ random distributed phasors. The random phasors represent the elementary and monochromatic disturbances to the true signal we aim to estimate. Since many noise sources can contribute to the disturbances (e.g. narrow-band TV transmitters, electronics noise, atmospheric disturbances, broadband galactic background, and so on), it is reasonable to assume $\mathcal{N}$ to be large. We make further assumptions on the noise phasors. We assume that the amplitude and phase characterizing them are statistically independent. Furthermore, we assume that their amplitudes are identically distributed and that their phases are uniformly distributed between $[-\pi,\pi)$. According to these conditions, it can be demonstrated that the marginal probability density function (PDF) of the length of the resultant phasor $\Vec{n}$, given by the noise random phasors' sum, is Rayleigh\footnote{The Rayleigh distribution is a $\chi$ distribution with two degrees of freedom after the rescaling of the random variable by a constant factor.} distributed. According to \cite{StatOptics}, we can express our measurement $\Vec{m}$ as the sum of $\Vec{s}$ and $\Vec{n}$ (as depicted in figure \ref{fig:phasors}), whose amplitude $a$ follows the marginal PDF given by the Rice distribution:  
\begin{equation}\label{eq:rice_pa}
 p_\mathrm{a}(s,\sigma)=
    \begin{cases}
      \,\frac{a}{\sigma^2}\cdot \mathrm{exp} \left(  -\frac{a^2+s^2}{2\sigma^2}  \right)\cdot I_0\left(\frac{as}{\sigma^2}\right) & \text{a $>$ 0}\\
      \, 0 & \text{a $\leq$ 0}, \\
    \end{cases} 
\end{equation}
In eqn. \ref{eq:rice_pa}, $s$ is the amplitude of the signal, $\sigma$ is the scale parameter of the function and represents the noise level of the measurement, and $I_0$ is the modified Bessel function of the first kind and order zero. When the signal is large compared to the noise, the Rice distribution can be approximated with a normal distribution having a standard deviation equal to $\sigma$ and centered on $\sqrt{s^2+\sigma^2}$. For low signals, the Rice distribution approaches a Rayleigh distribution. In our case, each of the $M$ amplitudes measured in the signal window will follow a different Rice distribution depending on the unknown underlying signal and noise level of the frequency bin. Note that the described formalism is valid both in the time and frequency domains, thus the development of further applications is possible in both domains. We refer the reader to Appendix\ \ref{sec:s_ML_estimator} if interested in the method we have developed to estimate the amplitudes $s(\nu_j)$ and their errors based on maximizing the Rice likelihood function. In Appendix\ \ref{sec:s_ML_estimator}, we also describe how to estimate the noise level of the bin over the $N$ noise windows.

\begin{figure*}[h!]
\centering
  \includegraphics[width=0.91\linewidth]{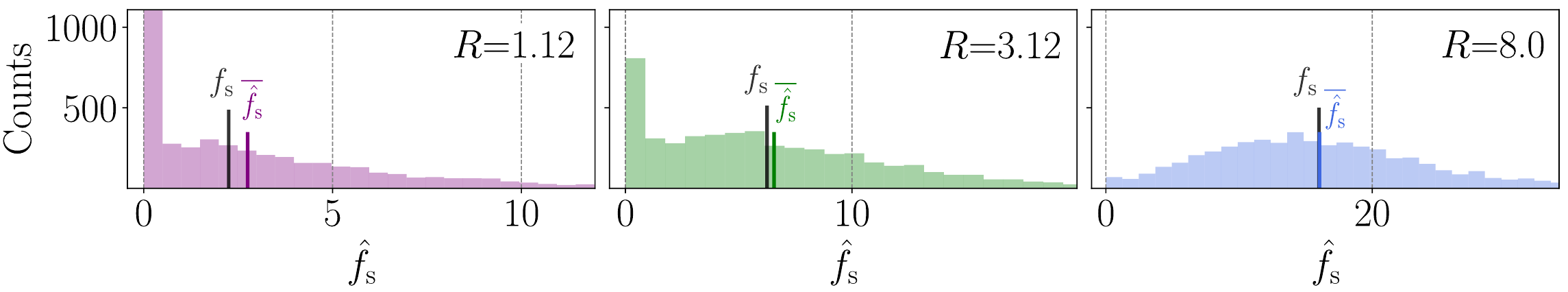}
\caption{Histograms of the signal fluence estimators obtained by running a toy Monte Carlo for three fixed values of $R$. The vertical black lines indicate the underlying true signal fluence used in the toy Monte Carlo. For comparison, the mean values of the histograms are shown through the colored vertical lines.}
\label{fig:hist_chi2_toy}
\end{figure*}
\subsection{Energy Fluence Estimator and its Uncertainty} 
\label{sec:fluence_nck2}

To evaluate the energy fluence in the frequency domain, we recall Parseval's theorem:
\begin{equation}
f_\mathrm{pol}=\epsilon_0c\Delta t \sum_{j=0}^{N_\mathrm{s}-1} E^2_\mathrm{pol}(t_j) \approx 2\,\epsilon_0c\Delta\nu \Delta t^2 \sum_{j=0}^{M-1}|D(v_j)|^2,
\label{eq:fluence_freq}
\end{equation}
where $N_\mathrm{s}$ is the number of samples considered belonging to the time trace, $D$ are the corresponding complex-valued DFTs,  $M$\,$=$\,$\frac{N_\mathrm{s}}{2}$ is the number of positively valued frequencies, $\Delta\nu$ is the frequency sample size, and the time sample size is given by $\Delta t$\,$=$\,$ \frac{1}{N_\mathrm{s} \Delta \nu}$. We express eqn. \ref{eq:fluence_freq}\footnote{In eqn.\ref{eq:fluence_freq}, we approximate $\epsilon_0c\Delta\nu \Delta t^2\big(|D(v_0)|^2+2\,\sum_{j=1}^{M-2}|D(v_j)|^2 +|D(v_{M-1})|^2 \big)$ by double counting the contribution from the first and the last bin, which are typically zero due to bandpass filtering.} as a sum of the fluences for each frequency:
\begin{equation}
f(\nu_j)=K\,|D(v_j)|^2  \,\, \rightarrow \,\, f_\mathrm{pol}=\sum_{j=0}^{M-1}f(\nu_j),
\label{eq:fluence_freq_contr}
\end{equation} 
where $K$\,=\,$2\,\epsilon_0\,c\, \Delta \nu \, \Delta t^2$.  Let us now consider the ideal case of a signal in the complete absence of noise. The energy fluence of the unknown signal is:
\begin{equation}
f_\mathrm{s}=K \sum_{j=0}^{M-1}s^2(v_j)=\sum_{j=0}^{M-1}f_\mathrm{s}(\nu_j).  \label{eq_fs}
\end{equation}
The estimator of the polarisation fluence and the estimator of the total fluence at the antenna position will be of the form:
\begin{equation}
\hat{f}_\mathrm{pol} =\sum_{j=0}^{M-1}\hat{f}_\mathrm{s}(\nu_j) \,\, \rightarrow \,\,
\hat{f}_\mathrm{tot} = \sum_\mathrm{pol} \hat{f}_\mathrm{pol},
\end{equation}
where $\hat{f}_\mathrm{s}(\nu_j)$ are the signal fluence estimators in the \text{$j$-th} frequency. The fluence of the same signal measured in the presence of random noise will be:
\begin{equation}
\hat{f}_\mathrm{a}=K \sum_{j=0}^{M-1}a^2(v_j)=\sum_{j=0}^{M-1}\hat{f}_\mathrm{a}(\nu_j), 
\label{eq_fa}
\end{equation}
with $\hat{f}_\mathrm{a}(\nu_j)$ being the frequency-dependent contributions to the amplitude fluence measured in the signal window. \\

For simplicity, let us analyze a single-frequency bin. We estimate the noise fluence of the \text{$j$-th} frequency by exploiting the $N$ noise windows: 
\begin{equation}
\hat{f}_\mathrm{n}(\nu_j)=\frac{K}{N}\sum_{i=0}^{N-1} n_i^2(v_j),
\label{eq_fn}
\end{equation}
where $n_i(v_j)$ indicates the Rayleigh-distributed random noise of the \text{$j$-th} bin measured in the \text{$i$-th} noise window. In eqn.\ref{eq_fn}, we use the sample mean over the windows, but depending on the characteristic noise of the measurements, one could consider excluding some of the noise windows from the evaluation, or using a more robust estimator as the one of eqn. \ref{eq:fn_median}. Finally, we define the fluence estimator of the signal in the \text{$j$-th} frequency by subtracting the noise fluence from the amplitude fluence estimators: 
\begin{equation} 
\hat{f}_\mathrm{s}(\nu_j)=
    \begin{cases}
      \hat{f}_\mathrm{a}(\nu_j)-\hat{f}_\mathrm{n}(\nu_j) & \text{$\hat{f}_\mathrm{a}(\nu_j)\geq\hat{f}_\mathrm{n}(\nu_j)$}\\
      \, 0 & \text{$\hat{f}_\mathrm{a}(\nu_j) < \hat{f}_\mathrm{n}(\nu_j)$}. \\
    \end{cases}   
\label{eq:fluence_nchi2}
\end{equation}
The statistical uncertainty can be estimated as: 
\begin{equation} 
\delta(\hat{f}_\mathrm{s}(\nu_j)) = \sqrt{\hat{f}_\mathrm{n}(\nu_j)\,\big(\hat{f}_\mathrm{n}(\nu_j)\,+2\,\hat{f}_\mathrm{s}(\nu_j)\big)}
\label{eq:err}
\end{equation}
Finally, the fluence estimator of the polarisation and the total fluence estimator can be calculated, as well as their uncertainties.
\begin{figure*}[h!]
\centering
\includegraphics[width=0.86\linewidth]{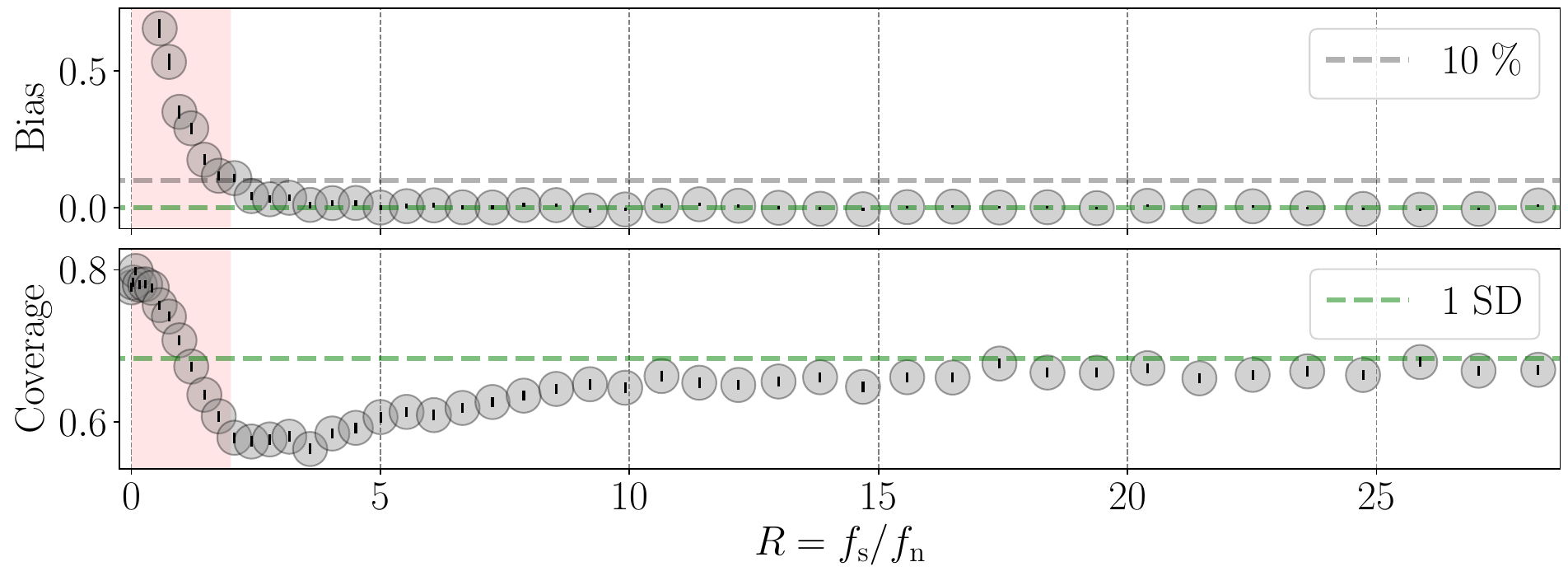}
\caption{Intrinsic bias of the signal fluence estimator (top) and errors coverage (bottom) obtained by running a toy Monte Carlo for several values of $R$. The orange shadowed area indicates the $R$ range which is strongly biased. In both plots, the error bars are divided by $\sqrt{N_\mathrm{bin}}$, with $N_\mathrm{bin}$ being the number of entries in the bin. }
\label{fig:bias_f}
\end{figure*}
\subsection{Toy Monte Carlo: Bias and Error Coverage for Single-frequency Estimator}
\label{sec:toy}
In eqn.\ref{eq:fluence_nchi2}, we included the second condition to prevent the signal estimator of a frequency bin from assuming negative unphysical values. , that we evaluated with a toy Monte Carlo. Exploiting the fact that the random variables $a$ are Rice-distributed, we generate $N_{\mathrm{MC}}$\,$=$\,5,000 random variables by setting $s$, and thus $f_\mathrm{s}$, to fixed values. Without loss of generality, we keep $\sigma$\,$=$\,1 for simplicity and we define the noise parameter $f_\mathrm{n}$\,$:=$\,2\,$\sigma^2$. For each amplitude $a$, we evaluate the noise fluence estimator $\hat{f}_\mathrm{n}$ as the mean over $N$ random variables following a normal distribution having mean $\mu$\,$=$\,2\,$\sigma^2$ and variance Var$\,=\,$4\,$\sigma^4/N$, where $N$ is the number of noise windows used in this work (see Appendix \ref{sec:derivation_fluence_nck2} for a detailed explanation). Finally, we evaluate the signal fluence estimator $\hat{f}_\mathrm{s}$ and its uncertainty. \\

In figure \ref{fig:hist_chi2_toy}, we show some examples of estimator distributions generated with the toy Monte Carlo for three fixed values of $R$\,$=$\,1.12, 3.12, 8.0, the signal-to-noise ratio of the frequency bin given by $R$\,$=$\,$f_\mathrm{s}/f_\mathrm{n}$. In the histograms relative to $R$\,$=$\,1.12, 3.12, the presence of a peak in correspondence of the bin including the estimators $\hat{f}_\mathrm{s}$\,$=$\,0 can be noticed. On the contrary, in the histogram relative to $R$\,$=$\,8.0, the number of entries in the same bin decreases significantly. This reflects a bias that gets less prominent with increasing values of $R$. We tackled the dependence on $R$ by scanning several values and computing the relative intrinsic bias and its standard deviation as:
\begin{equation}
\psi=\overline{\hat{f}_\mathrm{s}}/f_\mathrm{s} - 1, \,\,\, \sigma_\psi=\sqrt{\sum_{i=0}^{N_\mathrm{MC}-1}\frac{(\hat{f}_{\mathrm{s},i} -\overline{\hat{f}_\mathrm{s}})^2}{N_\mathrm{MC}-1}},  
\end{equation}
where $\overline{\hat{f}_\mathrm{s}}$ is the mean value of the estimators obtained through the toy Monte Carlo. As shown in the upper plot of figure \ref{fig:bias_f}, the relative bias is larger than 10\% up to $R$\,$=$\,2. For higher values of $R$, the bias decreases and can be neglected. The bias of the single-frequency estimator seems to affect the bias of the fluence estimator of a single polarisation mostly up to SNR$\,\approx\,$2.5 (see sec.\ \ref{trace_results}). We studied also the coverage of the errors as a function of $R$. We define the coverage as the percentage of data satisfying the condition:
\begin{equation}
f_\mathrm{s} \in
[\hat{f}_\mathrm{s} - \delta(\hat{f}_\mathrm{s}),\hat{f}_\mathrm{s}  + \delta(\hat{f}_\mathrm{s})].   
\end{equation}
As shown in the lower plot of figure \ref{fig:bias_f}, in correspondence of large bias, we overestimate the errors, while for $R$>2 the coverage approaches the classical definition of standard deviation. 
\begin{figure*}[h!]
\centering
\includegraphics[width=0.94\linewidth]{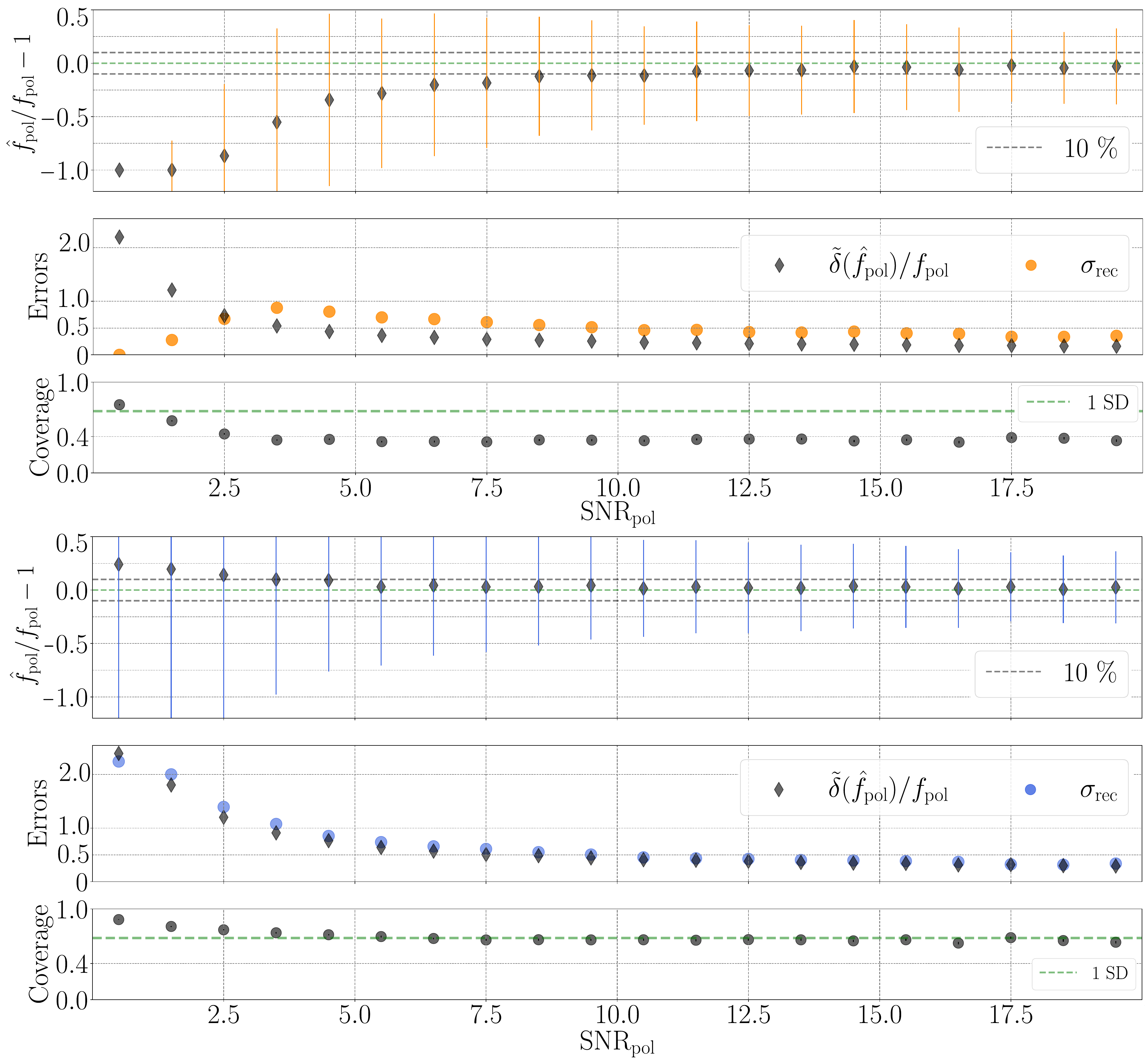}
\caption{Estimation of $f_\mathrm{pol}$ for the noise subtraction method (upper figure) and the Rice-distribution method (lower figure) up to SNR$_\mathrm{pol}$\,=\,20 (as defined in eqn. \ref{eq:snr_trace}). In the upper plots, we show the bias together with the error bar given by $\sigma_\mathrm{rec}$. In the middle plots, the resolution of each bin is compared with $\tilde{\delta}(\hat{f})/f$ of the same bin on a logarithmic scale. The coverage of the uncertainties (eqn. \ref{eq:coverage}), together with its error normalized to the number of entries of the bins, are shown in the lower plots.}
\label{fig:both_trace_up20}
\end{figure*}
\section{Validation of the Method} 
\label{Analysis}
In this section, we validate the Rice-distribution method and compare it to the noise subtraction method. Both methods are applied to simulated noisy electric field traces (sec.\ \ref{efield_meas}). No additional signal cleaning or RFI suppression is applied. The fluence estimators are compared to the reference values evaluated from the same electric fields simulated in the absence of noise (sec.\ \ref{efield_sim}). In this way, the analysis is not affected by biases potentially introduced by the unfolding of the antenna response. Since the methods differ, we compare the estimators to different reference values. Concerning the noise subtraction method, the reference values are calculated in the time domain by evaluating eqn.\ \ref{eq:poynting_component} in the signal window. Because of the usage of windowing in the Rice-distribution method, this equation would yield a different value\footnote{For the window function adopted, we estimate an average difference of 2\% for antennas within 1.5 Cherenkov radii.}. Instead, the reference values are obtained evaluating eqn.\ \ref{eq:fluence_freq_contr} over the frequency spectrum of the signal window after applying the Tukey window and excluding the frequencies outside the 30-80\,MHz band. In the following, error coverage, bias, and uncertainty for both methods are analyzed as functions of the SNR. 

For a single polarisation, we define the signal-to-noise ratio as: 
\begin{equation}
\mathrm{SNR}_\mathrm{pol}=\bigg(\frac{|E_\mathrm{pol}^\mathrm{Hilb}(t_\mathrm{peak})|}{\mathrm{RMS}_\mathrm{pol}}\bigg)^2,
\label{eq:snr_trace}
\end{equation}
where $t_\mathrm{peak}$ is the estimated pulse position, $E_\mathrm{pol}^\mathrm{Hilb}(t_\mathrm{peak})$ is the amplitude of the Hilbert envelope of the electric-field component at the pulse position, and $\mathrm{RMS}_\mathrm{pol}$ is the root-mean-square of the Hilbert envelope in the noise window. Similarly, we define the signal-to-noise ratio over the three polarisations as:
\begin{equation}
\mathrm{SNR}_\mathrm{tot}=\bigg(\frac{A_\mathrm{tot}^{\mathrm{Hilb}}(t_\mathrm{peak})}{\mathrm{RMS}_\mathrm{tot}}\bigg)^2, 
\label{eq:snr_station}
\end{equation}
with $\mathrm{RMS}_\mathrm{tot}$ being the root-mean-square in the noise window of the trace:
\begin{equation}
A_\mathrm{tot}^{\mathrm{Hilb}}(t) = \sqrt{\sum_\mathrm{pol} \big|E_\mathrm{pol}^\mathrm{Hilb}(t)\big|^2}. 
\label{eq:hilbert_mag}
\end{equation}
In real data, the position of pulses with a small signal-to-noise-ratio is not known. It can, however, be estimated, for example by fitting a radio wavefront model \cite{wavefront} to pulses with sufficient signal-to-noise ratio. In this work, we assume that the approximate pulse position is known. To this end, we start from the Monte Carlo position $t_\mathrm{peak}^\mathrm{MC}$ and then determine $t_\mathrm{peak}$ by evaluating the maximum of the Hilbert envelope around $t_\mathrm{peak}^\mathrm{MC}$\,$\pm$30\,ns. Simulated traces strongly affected by thinning artifacts are excluded by applying a cut based on antenna position. Following \cite{Schluter:2022mhq}, we exclude antennas with a distance from the shower axis greater than 2 Cherenkov radii.
\begin{figure*}[h!]
\centering
\includegraphics[width=0.94\linewidth]{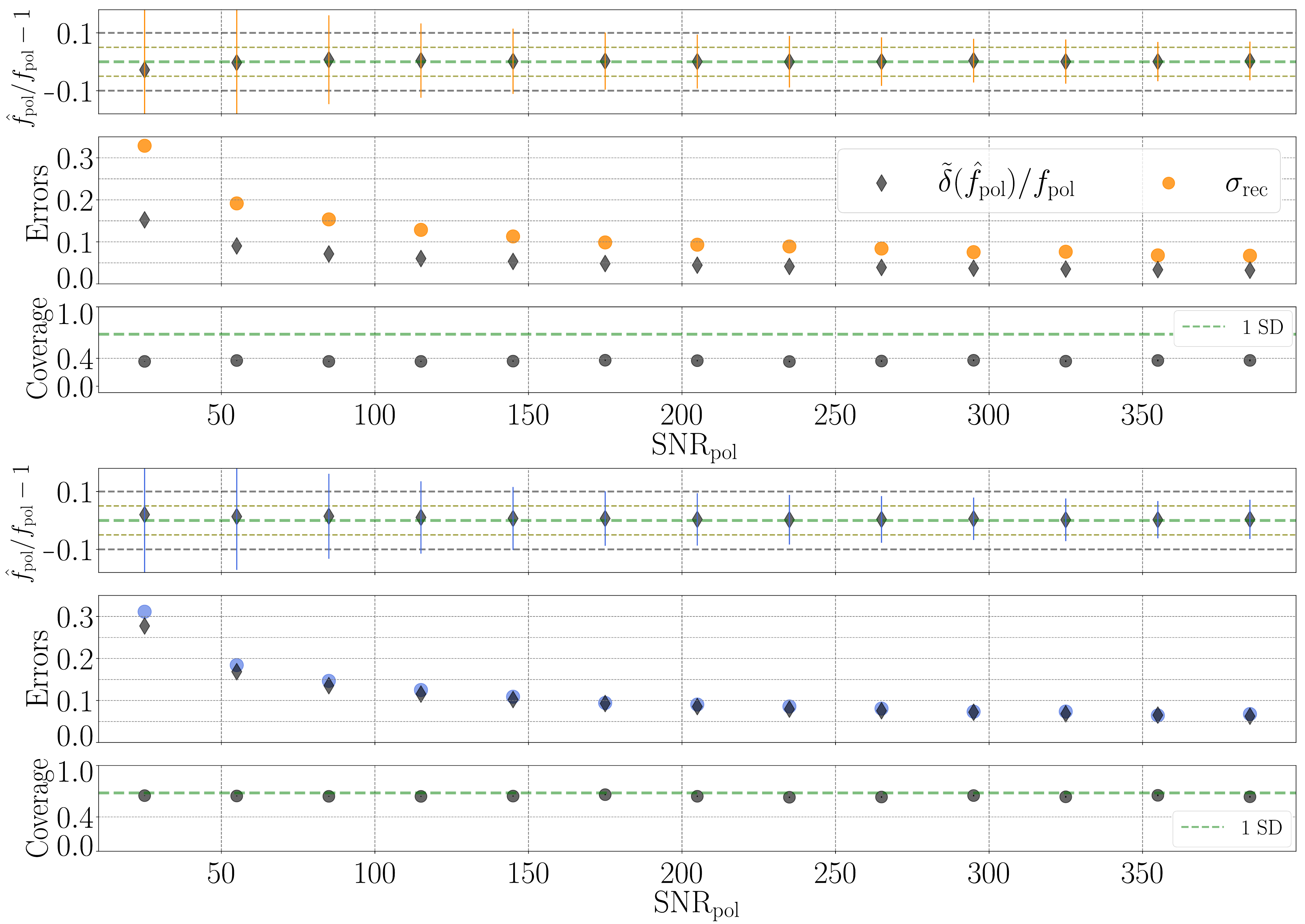}
\caption{Estimation of $f_\mathrm{pol}$ for the noise subtraction method (upper figure) and the Rice-distribution method (lower figure) up to up to SNR$_\mathrm{pol}$\,=\,400 (as defined in eqn. \ref{eq:snr_trace}).}
\label{fig:both_trace_up400}
\end{figure*}
\subsection{Polarisation Fluence $f_\mathrm{pol}$}\label{trace_results} 
For each electric field component, we evaluate the fluence estimators $\hat{f}_\mathrm{pol}$ and their uncertainties $\delta(\hat{f}_\mathrm{pol})$\footnote{We used the estimator of eqn. \ref{eq:fn_median} to estimate $\hat{f_n}(\nu_j)$, the noise fluence of the $j$-th frequency bin.}, along with the reference values $f_\mathrm{pol}$ and the signal-to-noise ratios $\mathrm{SNR}_\mathrm{pol}$. After calculating the ratios $\psi$\,=\,$\frac{\hat{f}_\mathrm{pol}}{f_\mathrm{pol}}$, we examine different signal-to-noise ratio intervals. Electric fields are broken down into East-West (EW), North-South (NS), and Vertical (V) components. We combine the data irrespective of the polarisation, collecting all ratios $\psi$ within a given SNR range without distinguishing between EW, NS, and V components. To reduce the impact of outliers, in each SNR interval, we calculate $\widetilde{\psi}$, the median value of $\psi$. Finally, we define the average reconstruction bias in the bin as $\mathrm{Bias}$\,$=$\,$\widetilde{\psi}-1$. The dispersion of each bin is described by the interquartile range (i.q.r.) of the distribution. We calculate the equivalent standard deviation as $\sigma_\mathrm{rec}$\,$\approx$\,$\frac{\mathrm{i.q.r.}}{1.35}$. 

As shown in figure \ref{fig:both_trace_up20}, where SNR values up to 20 are investigated, we found that for the noise subtraction method, SNR values below 2.5 are characterized by an extremely large bias (exceeding 75\%). In the first bin, the bias reaches 100\%, as the negative-valued estimators have been set to zero, significantly influencing the result. For larger values, the bias decreases: starting from SNR\,$\approx$\,9, the average bias is contained within 10\%. In many works, SNR values of at least 10 are required for higher-level analyses \cite{xmx_PRD, MarvinARENA2024, MaxEnergyScaleARENA2024}. In comparison, the Rice-distribution method exhibits a smaller bias. On average, up to SNR\,$\approx$\,2.5, the bias fluctuates between 25\% and 15\%. The bias in this region is probably due to a combination of the estimator intrinsic bias discussed in sec.\ \ref{sec:toy}, and the lack of RFI suppression. For increasing SNR values, the bias fluctuates between 10 and 5\%, and can be neglected.    
We now focus on the uncertainty estimation of the two methods. We compare $\tilde{\delta}(\hat{f}_\mathrm{pol})/f_\mathrm{pol}$, the median value of the relative uncertainties of each bin, with $\sigma_\mathrm{rec}$. For unbiased estimators, we would expect the uncertainties to reflect the reconstruction resolution. The uncertainties of the noise subtraction method result in being systematically underestimated, as shown in figure \ref{fig:both_trace_up20}. To further corroborate this conclusion, we investigate the uncertainty coverage of the method. We define the coverage as the percentage of data points satisfying the condition:
\begin{equation}
f_\mathrm{pol}\in
[\hat{f}_\mathrm{pol} - \delta(\hat{f}_\mathrm{pol}),\hat{f}_\mathrm{pol} + \delta(\hat{f}_\mathrm{pol})].  
\label{eq:coverage}
\end{equation}
The error coverage of the noise subtraction method oscillates between 30\% and 40\%. The coverage of the  Rice-distribution method aligns with its bias. For SNR values above 2.5, where the bias is less significant, the coverage fluctuates around 68\%, close to the classical definition of one standard deviation (SD) coverage.
 
We found that for higher values of SNR, both methods converge to a reconstruction bias below 5\%, as shown in figure \ref{fig:both_trace_up400}, where the region up to SNR$\,=\,$400 is studied. For a fair comparison, in all of the plots of the figure, we exclude the data points corresponding to SNR\,$<$\,10, where the noise subtraction method is strongly biased. Once more, the noise subtraction method systematically underestimates the uncertainties. The average coverage is about 40\%, while our method provides a more consistent way of uncertainty estimation, with an error coverage slightly below the desired 68\%. 
\begin{figure*}[b!]
\centering
\includegraphics[width=0.94\linewidth]{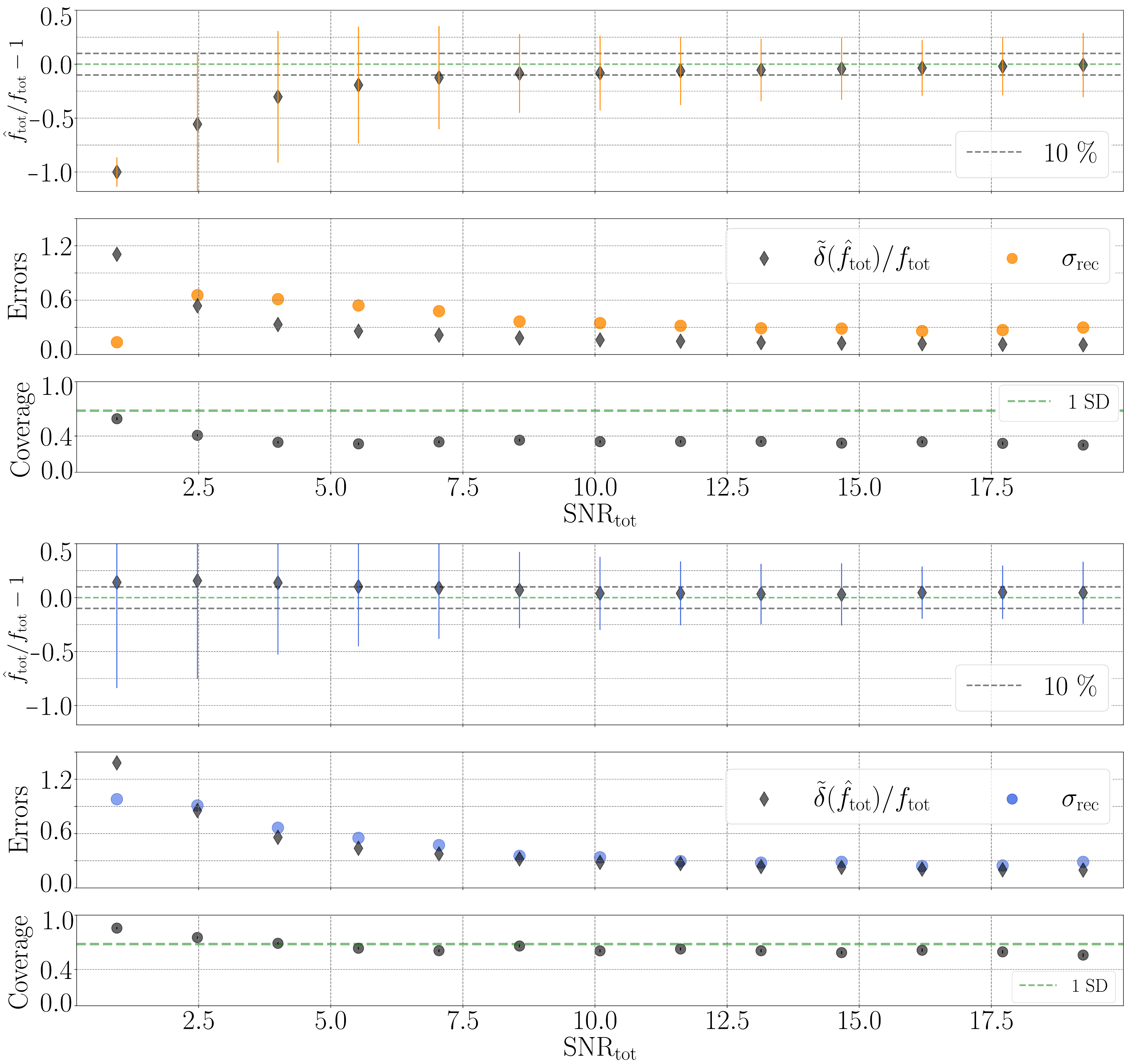}
\caption{Estimation of $f_\mathrm{tot}$ for the noise subtraction method (upper figure) and the Rice-distribution method (lower figure) up to SNR$_\mathrm{tot}$\,$=$\,20 (as defined in eqn. \ref{eq:snr_station}).}
\label{fig:both_station_up20}
\end{figure*}
\subsection{Total Fluence $f_\mathrm{tot}$}\label{antenna_results}
\begin{figure*}[h!]
\centering
\includegraphics[width=0.94\linewidth]{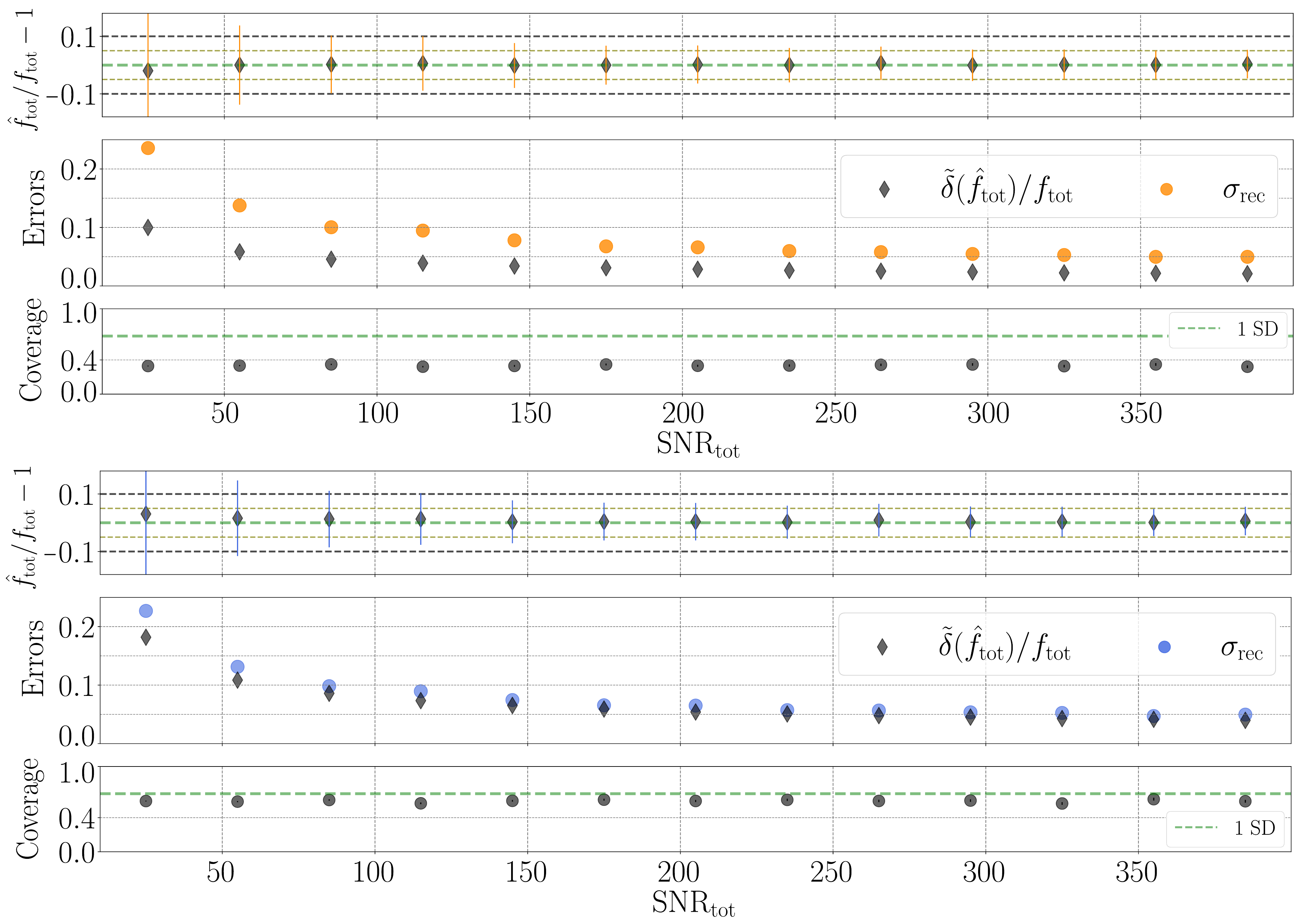}
\caption{Estimation of $f_\mathrm{tot}$ for the noise subtraction method (upper figure) and the Rice-distribution method (lower figure) up to SNR$_\mathrm{tot}$\,$=$\,400 (as defined in eqn. \ref{eq:snr_station}).}
\label{fig:both_station_up400}
\end{figure*}

We repeat the analysis presented in the previous section, adapting it to the estimation of the total fluence evaluated at the antenna position. The noise subtraction method exhibits a large bias at the lowest SNR values, still exceeding 10\% up to SNR\,$\approx$\,6.5. From this value, the bias starts fluctuating around 10\% and then stabilizes within 5\% with increasing SNR (see figure \ref{fig:both_station_up20}). The uncertainties derived from the method are underestimated, with 
an average coverage of about 35\%. We can draw the same conclusion for higher values of SNR, despite the bias remaining contained within 5\% (see figure \ref{fig:both_station_up400}, where we show the SNR range from 10 up to 400). In contrast, the bias of the method we implemented fluctuates around 10\% up to SNR\,$\approx$\,7.5, and around 5\% with increasing SNR (see figures \ref{fig:both_station_up20}, \ref{fig:both_station_up400}). On average, the error coverage fluctuates around 60\% at any SNR value.
\section{Discussion}
In this article, we focused on an unbiased comparison of signal-estimation methods, using a generic approach and not tailoring their application to any specific experiment. The Rice distribution method shows promising improvements, especially for incorporating low-SNR signals. The extent of improvement achievable in the reconstruction of shower observables, such as primary particle energy or depth of shower maximum, will be dependent on the specifics of said reconstruction algorithm. For example, it may be limited by the reliability of the lateral distribution function at low SNR values. Other technical aspects should also be considered in higher-level analyses. For example, due to the power loss associated with windowing, careful selection of the window function, length, and tapering percentage is necessary. This step may require renormalizing fluences and radiation energies when relating them to particle energies. Lastly, if pulse position determination is inaccurate, it could introduce additional biases in the fluence estimation. We note that none of these limitations are specific to the Rice method for signal estimation, which will yield a minimally biased estimate of the energy fluence and an adequate quantification of its uncertainty. The Rice method thus provides the potential to incorporate data from lower-SNR measurements in reconstruction procedures based on fluence estimation, when proper care is taken regarding limitations such as the ones described above.

\section{Conclusions}
In this work, we quantified the bias at low SNR values in the noise-subtraction method, which is conventionally used to reconstruct the energy fluence. We also quantified the underestimation of the uncertainties associated with this method. To address these known limitations, we developed a method based on Rice distributions for a more accurate estimation of the fluence at low SNR values and a correct evaluation of its uncertainty. At the antenna level, we achieve an average bias of 10\% up to signal-to-noise ratios of about 7.5. Above this value, the bias fluctuates around 5\%. We also significantly improved the estimation of the uncertainties, reaching a reliable coverage close to 68\%. With the Rice-distribution method, the estimation of energy fluence and its uncertainty can thus be significantly improved, and signal-to-noise cuts can be lowered or possibly completely avoided in higher-level analyses. 

The Rice-distribution method is generic and can be applied to any kind of radio observations of air showers and in-ice cascades. Since the method is not constrained by the frequency bandwidth of sensitivity, radio experiments in the field of astroparticle physics can make use of it. Among others, the Pierre Auger Observatory, RNO-G \cite{RNOG}, LOFAR, ANITA, GRAND \cite{GRAND}, SKA \cite{SKAlow}, and the IceCube Neutrino Observatory \cite{Icecube_radio} could benefit from adopting this approach. The statistical background employed to develop the fluence estimation method can be further exploited, as we show in the Appendix \ref{sec:s_ML_estimator}. There, we describe how to estimate the signal amplitude employing the Rice likelihood function. Further improvements and applications are still feasible by combining this statistical approach and the signal modeling in the frequency domain. 

\section*{Acknowledgements}
The simulations used in this work were processed through $\overline{\mathrm{Off}}\underline{\mathrm{line}}$, the reconstruction framework of the Pierre Auger collaboration. We thank the collaboration for allowing us to use the detector simulation implemented within the framework and for providing us access to their radio background measurements. We would also like to thank our colleagues in the Pierre Auger Collaboration for fruitful discussions and helpful input. We thank the Hermann von Helmholtz-Gemeinschaft Deutscher Forschungszentren e.V. for supporting Sara Martinelli through the  Helmholtz International Research School for Astroparticle Physics (HIRSAP), part of \textit{Initiative and Networking Fund} (grant number HIRS-0009).

\appendix
\label{appendix}
\section{Derivation of the Fluence Estimator and its Uncertainty} 
\label{sec:derivation_fluence_nck2}

We derive the energy fluence estimator and its uncertainty as presented in sec.\ \ref{sec:fluence_nck2}. For a monochromatic signal with \( K=1 \), we rewrite equations \ref{eq_fs}, \ref{eq_fa} as:
\begin{equation}
f_s = s^2(\nu_0) \equiv s^2, \quad \hat{f}_\mathrm{a} = a^2(\nu_0) \equiv a^2.
\end{equation}
We aim for an estimator of \( f_s \) proportional to \( a^2 \). We standardize \( a \) as $b\,$$=$\,$a/\sigma$, where \( b \sim p_\mathrm{a}(s/\sigma, 1) \), equivalent to a non-central \(\chi\)-distribution with two degrees of freedom (DF) and non-centrality parameter $\lambda\,$$=$\,$s/\sigma$.
Thus, \( b^2 \) follows a non-central \( \chi^2 \)-distribution:
\begin{equation}
b^2 \sim \chi^2_\mathrm{nc} (DF=2, \lambda=(s/\sigma)^2),
\end{equation}
with $E(b^2)$\,$=$\,$2+(s/\sigma)^2$ and $\mathrm{Var}(b^2)$\,$=$\,$2\,(2+2\,(s/\sigma)^2)$.
\subsubsection*{Noise Fluence Estimator}
Estimating the signal fluence requires a noise fluence estimator. We sample the noise trace in $N$ windows, each yielding a noise amplitude \( n_i \) assumed to be Rayleigh-distributed. Each \( (n_i/\sigma)^2 \) follows a chi-square distribution with two DF, so the sum \( T = \sum_{i=0}^{N-1}(n_i/\sigma)^2 \) follows a chi-square distribution with \( 2N \) DF, approximated by a normal distribution with mean \( \mu = 2N \) and variance \( \mathrm{Var} = 4N \) for large \( N \). The noise fluence estimator is the sample mean of the squared noise amplitudes:
\begin{equation}
\hat{f}_\mathrm{n} = \frac{1}{N} \sum_{i=0}^{N-1} n_i^2 = \frac{\sigma^2}{N} T,
\end{equation}
which is normally distributed with mean $\mu\,$$=$\,$2\sigma^2$ and variance $\mathrm{Var}\,$$=$\,$\sigma^4/N$. Defining $f_\mathrm{n}$\,$:=$\,$2\sigma^2$, we find \( \hat{f}_\mathrm{n} \) to be an unbiased estimator of \( f_\mathrm{n} \). For $N$\,$\approx$\,60, its relative uncertainty is around 13\%.
We derive a more robust estimator in the presence of outliers using the population median, as:
\begin{equation}
\hat{f}_\mathrm{n} = \frac{2}{1.405}\cdot\text{median}\big[n_i^2\big],
\label{eq:fn_median}
\end{equation}
where we exploit that the median value of a $\chi^2_\mathrm{DF}$ distribution can be approximated as DF$\cdot\big(1-\frac{2}{9\cdot\,\mathrm{DF}}\big)^3$, and that $n_i^2$\,$\sim$\,$\sigma^2\cdot\chi^2_\mathrm{2}$.
\subsubsection*{Signal Fluence Estimator}
We derive the expected value of $\hat{f}_\mathrm{a}$ as: 
\begin{equation} 
E(\hat{f}_\mathrm{a})=E(a^2)=\sigma^2\,E(b^2)=2\,\sigma^2 + s^2
\end{equation}
By setting $s$\,$=$\,0, for the noise estimator we get:  
\begin{equation} 
E(\hat{f}_\mathrm{n})= 2\,\sigma^2:=f_\mathrm{n}.
\end{equation}
Estimating the signal fluence as $\hat{f}_\mathrm{s}$\,$=$\,$\hat{f}_\mathrm{a}-\hat{f}_\mathrm{n}$, its mean value will be:
\begin{equation} 
E(\hat{f}_\mathrm{s})=E(\hat{f}_\mathrm{a})-E(\hat{f}_\mathrm{n})=s^2=f_\mathrm{s},
\end{equation}
meaning that such an estimator would be unbiased. Since the energy fluence has to be positively defined, we have to introduce the condition:
\begin{equation} 
\hat{f}_\mathrm{s}=0 \quad \hat{f}_\mathrm{a}<\hat{f}_\mathrm{n}.
\label{limit}
\end{equation}
As discussed in sec.\ \ref{sec:toy}, the above condition introduces a positive bias that can be neglected for $R$\,$>$\,2. 

\subsubsection*{Signal Fluence Uncertainty}
We derive the variance of $a^2$ : 
\begin{equation} 
\mathrm{Var}(a^2)=\sigma^4\,\mathrm{Var}(b^2)=2\,\sigma^2(2\,\sigma^2+2\,s^2).  
\end{equation}
To derive the variance of $\hat{f}_\mathrm{s}$, we make the assumptions:
\begin{itemize}
  \item The noise fluence is measured with much better precision than $a^2$. This can be considered a valid assumption since the variable $a$ is a single measurement (from the signal window), while the noise fluence is estimated through $N$\,$\approx$\,60 noise windows; 
  \item The probability of hitting the physical limit of eqn.\ref{limit} is negligible, i.e. we are assuming $R$\,$>$\,$2$; 
  \item $(s/\sigma^2)$ is large, meaning that $a^2$ and $b^2$ are normal in a good approximation, and, again $R$ is large.
\end{itemize}
Under these assumptions, the variance of $\hat{f}_\mathrm{s}$ is:
\begin{equation} 
\mathrm{Var}(\hat{f}_\mathrm{s})=\mathrm{Var}(a^2)\approx \hat{f}_\mathrm{n}\big(\hat{f}_\mathrm{n}+2\,\hat{f}_\mathrm{s}\big), 
\end{equation}
where we approximate the parameters as $s^2$\,$\approx$\,$\hat{f}_\mathrm{s}$ and $2\,\sigma^2$\,$\approx$\,$ \hat{f}_\mathrm{n}$. We define the confidence interval as $\hat{f}_\mathrm{s}$\,$\pm$\,$\delta(\hat{f}_\mathrm{s})$, with:  
\begin{equation} 
\delta(\hat{f}_\mathrm{s})=\sqrt{\mathrm{Var}(\hat{f}_\mathrm{s})}=\sqrt{\hat{f}_\mathrm{n}\big(\hat{f}_\mathrm{n}+2\,\hat{f}_\mathrm{s}\big)}. 
\end{equation}

\section{Spectral Amplitude Estimator and its Uncertainty} 
\label{sec:s_ML_estimator}

The theoretical background described in sec.\ \ref{sec:rice_theo} has much more potential to exploit than what has been shown so far. Here we introduce the Rice likelihood function and provide an application example. In particular, we present a method to estimate the signal spectral amplitude and its uncertainty.

Let us consider the case of a single-frequency signal $s_0$\,$\equiv$\,$s(\nu_0)$, with $a_0$\,$\equiv$\,$a(\nu_0)$ being the measured spectral amplitude, and $\hat{\sigma}_0$ the estimator of the noise level. The Rice likelihood function will be:
\begin{equation}
L(s \mid a=a_0, \sigma=\hat{\sigma}_0) =  \frac{a}{\sigma^2}\, \mathrm{exp} \left(  -\frac{a^2+s^2}{2\sigma^2}  \right)I_0\left(\frac{as}{\sigma^2}\right), 
\end{equation}
where $s$\,$\ge$\,$0$. We evaluate $\hat{\sigma}_0$ over the $N$ noise windows:
\begin{equation}
\hat{\mu}_0=\frac{1}{N}\sum_{i=0}^{N-1} a_i(v_0) \,\, \rightarrow \,\, \hat{\sigma}_0\equiv\mu_0 \sqrt{2/\pi},
\end{equation}
where in the last equivalence we assumed the noise to be Rayleigh-distributed with scale parameter $\sigma_0$ and mean value $\mu_0$. We define the estimator of the parameter $s_0$ as the amplitude $\hat{s}_\mathrm{ML}$ that maximizes $L(s)$. Since it is not possible to solve analytically $\frac{dL(s)}{ds}\Bigr|_{\substack{s=\hat{s}_{\mathrm{ML}}}}$\,$=$\,0, we recommend the reader to run a scalar minimization algorithm to find the minimum of the \textit{cost} function $J(s)$\,$=$\,$-\mathrm{ln\big(}L(s)\mathrm{\big)}$. To avoid unphysical solutions, such as negative results, and to make the minimization process reliable, we exploit a bounded solver. We set as bounds $[a_0 - 1.5\,\hat{\sigma}_0$,\, $a_0 + 1.5\,\hat{\sigma}_0]$\footnote{The bounds were found by studying the first and second derivative spaces of the likelihood function.}. There are some intervals where it is not strictly necessary to run the minimizer solver. In particular, for $a_0/\hat{\sigma}_0 $\,$\leq$\,$ \sqrt{2}$, $\hat{s}_\mathrm{ML} = 0$. For larger ratios, as $a_0/\hat{\sigma}_0$\,$\geq$\,4, we can approximate $L(s)$ by a normal distribution having SD$\,=$\,$\hat{\sigma}_0$ and centered in $\sqrt{a_0^2-\hat{\sigma}_0^2}$, thus $\hat{s}_\mathrm{ML}$\,$\approx$\,$\sqrt{a_0^2-\hat{\sigma}_0^2}$. To evaluate the estimator uncertainty $\delta$, we work in Gaussian approximation. First, we approximate the likelihood $cost$ function as:
\begin{equation}
\begin{split}
 J(s)\approx  J(\hat{s}_{\mathrm{ML}}) + (s-\hat{s}_{\mathrm{ML}}) \cdot \frac{\partial J(s)}{\partial s} \Bigr|_{\substack{s=\hat{s}_{\mathrm{ML}}}}+\\  
  +(s-\hat{s}_{\mathrm{ML}})^2\cdot\frac{1}{2}\frac{\partial^2 J(s)}{\partial s^2}\Bigr|_{\substack{s=\hat{s}_{\mathrm{ML}}}},
\end{split}
\end{equation}
where the first two terms are null by definition. Since the $cost$ function of a Gaussian distribution is a parabolic function, we can approximate $J(s)$ to a parabola as well. This would lead to the following system of equations:
\begin{equation}
    \begin{cases}
    \begin{split}
    J(s)= (s-\hat{s}_{\mathrm{ML}})^2\cdot\frac{1}{2}\frac{\partial^2 J(s)}{\partial s^2}\Bigr|_{\substack{s=\hat{s}_{\mathrm{ML}}}}\\
    = \delta^2\cdot\frac{1}{2}\frac{\partial^2 J(s)}{\partial s^2}\Bigr|_{\substack{s=\hat{s}_{\mathrm{ML}}}}       
    \end{split}\\
    J(s)\approx\,k^2 \quad \rightarrow \quad \delta=k\cdot SD \\
    \end{cases}
\label{eq:system_eq}
\end{equation}
Finally, by requiring a one-sigma interval, i.e. $\delta$\,$=$\,1\,SD, the solution of eqn. \ref{eq:system_eq} is given by:
\begin{equation}
 \delta=1\Bigg/\sqrt{\frac{1}{2}\frac{\partial^2 J(s)}{\partial s^2}\Bigr|_{\substack{s=\hat{s}_{\mathrm{ML}}}}}. 
 \label{eq:error_hess}
\end{equation}

\section{Table of Variables and Notation} \label{appendix:table}
 
{\renewcommand{\arraystretch}{1.4}
\begin{center}
\begin{tabular}{||c c ||} 
 \hline
  \multicolumn{2}{||c||} {\textbf{Generic Variables}}   \\ [0.5ex] 
 \hline\hline
 $E_\mathrm{pol}$ & \makecell{Electric field component in the\\ chosen coordinates system}    \\ 
 \hline
 $f_\mathrm{pol}$ & \makecell{Energy fluence relative to the considered\\ electric field component}   \\
 \hline
 $f_\mathrm{tot}$ & Total energy fluence at the antenna position   \\
 \hline
 SNR$_\mathrm{pol}$ & \makecell{Signal-to-noise ratio relative to\\ the considered electric field component}    \\ 
  \hline
 SNR$_\mathrm{tot}$ & \makecell{Signal-to-noise ratio calculated over all\\ the electric field components}   \\   
 \hline
 \multicolumn{2}{||c||} {\textbf{Notation}}     \\ [0.5ex] 
 \hline\hline
 $y$ & Unknown parameter (true value)  \\ 
 \hline
 $\hat{y}$ & Estimator of the parameter (or measurement)  \\ 
 \hline
 $ \delta\big(\hat{y}\big)$ & Uncertainty on the estimator (or measurement)  \\ 
 \hline
 $\overline{h}$ & Mean value of the variable   \\
 \hline
 $\tilde{h} $ &  Median value of the variable   \\ [0.5ex]
 \hline
 \hline
\multicolumn{2}{||c||} {\textbf{Rice-distribution Variables (fixed frequency or time)}} \\ [0.5ex] 
 \hline\hline
 $s$ &  Signal amplitude in the  absence of noise  \\ 
 \hline
 $a$ & \makecell{Measured amplitude of the signal in the\\ presence of noise }  \\ 
 \hline
 $f_\mathrm{s}, \hat{f}_\mathrm{s}$ & Signal energy fluence, its estimator  \\ 
 \hline
 $f_\mathrm{n}, \hat{f}_\mathrm{n}$ & Noise energy fluence, its estimator     \\ 
 \hline
 $\hat{f}_\mathrm{a}$ & \makecell{Signal energy fluence measured in the\\ presence of noise}  \\   
 \hline
\end{tabular}
\end{center}}

\bibliographystyle{elsarticle-num}
\bibliography{cas-refs.bib}
\end{document}